\documentclass[twocolumn]{aastex63}
\usepackage{graphicx}
\usepackage{dcolumn}
\usepackage{bm}
\usepackage{amsmath}
\usepackage{longtable}

\usepackage{array}
\begin{document}

\title{High-energy neutrino production from AGN Disk Transients Impacted by Circum-disk Medium}

\author{Zi-Hang Zhou}
\affiliation{Department of Astronomy, School of Physics, Huazhong University of Science and Technology, Wuhan 430074, China; \url{kaiwang@hust.edu.cn}}

\author[0000-0002-9195-4904]{Jin-Ping Zhu}
\affil{School of Physics and Astronomy, Monash University, Clayton, VIC 3800, Australia; \url{jin-ping.zhu@monash.edu}}
\affil{OzGrav: The ARC Centre of Excellence for Gravitational Wave Discovery, Clayton Victoria 3800, Australia}

\author[0000-0003-4976-4098]{Kai Wang}
\affiliation{Department of Astronomy, School of Physics, Huazhong University of Science and Technology, Wuhan 430074, China; \url{kaiwang@hust.edu.cn}}

\begin{abstract}

Various supernovae (SN), compact object coalescences, and tidal disruption events are widely believed to occur embedded in active galactic nuclei (AGN) accretion disks and generate detectable electromagnetic (EM) signals. We collectively refer to them as \emph{AGN disk transients}. The inelastic hadronuclear ($pp$) interactions between shock-accelerated cosmic rays and AGN disk materials shortly after the ejecta shock breaks out of the disk can produce high-energy neutrinos. However, the expected efficiency of neutrino production would decay rapidly by adopting a pure Gaussian density atmosphere profile applicable for stable gas-dominated disks. On the other hand, AGN outflows and disk winds are commonly found around AGN accretion disks. In this paper, we present that the circum-disk medium would further consume the shock kinetic energy to more efficiently produce high-energy neutrinos, especially for $\sim$\,TeV$-$PeV neutrinos that IceCube detects. Thanks to the existence of circum-disk medium, we find that the neutrino production will be enhanced significantly and make a much higher contribution to the diffuse neutrino background. Optimistically, $\sim20\%$ of the diffuse neutrino background can be contributed by AGN disk transients.

\end{abstract}


\section{Introduction}

It has been suggested that in AGN, the accretion disk surrounding the supermassive black hole (SMBH) may contain a large population of stars and compact objects, including white dwarfs (WDs), neutron stars (NSs), and stellar-mass BHs. On the one hand, AGN disks can capture stars and compact objects from nuclear star clusters \citep{syer1991,artymowicz1993,macleod2020,fabj2020}. On the other hand, AGN stars can form at the outer self-gravitating region of the disk via gravitational instability, migrate into inner orbit, and end up as compact objects \citep{kolykhalov1980,shlosman1989,goodman2004,wang2011,wang2012,dittmann2020}. AGN accretion disks provide a natural environment for embedded stars and compact objects to grow, to accrete materials, and to migrate within it \citep[e.g.,][]{mckernan2012,bellovary2016,wang2021accretion2,kaaz2021,perna2021b,pan2021}. There could be abundant stars and compact objects gathering in the inner part of the AGN disks, and more easily forming binary systems there due to the high encountered probability and the dynamical friction within the disk \citep[e.g.,][]{baruteau2011,bartos2017,yang2021micro}. Thus, explosions of massive stars and mergers/collisions between stars and compact objects can frequently occur there \citep[e.g.,][]{bartos2017,leigh2018,yang2019,mckernan2020black,tagawa2020,tagawa2021,zhu2021thermonuclear,grishin2021supernova,li2021,li2022,li2022b}.

Stars in AGN disks can spin up to critical rotation via accretion and undergo quasi-chemically homogeneous evolution to become compact stars \citep{dittmann2021,cantiello2021,jermyn2021}, the properties of which are similar to those of Wolf-Rayet progenitors of long gamma-ray bursts (GRBs) and superluminous SNe \citep{yoon2006,woosley2006GRB,zhang2018}. Thus, core-collapse AGN stars are very likely to drive lGRBs and powerful SNe. Mergers and collisions between stars and compact objects in AGN disks potentially result in different kinds of transient phenomena, such as (1) Type Ia SNe (SNe Ia) from binary WDs and accretion-induced collapses of WDs \citep[e.g.,][]{whelan1973,nomoto1982}; (2) short GRBs and kilonovae from neutron star mergers \citep[e.g.,][]{eichler1989,narayan1992, li1998,metzger2010}; (3) tidal disruption events by stellar-mass BHs \citep[e.g.,][]{2020stone,yang2022}; etc. However, the surrounding dense atmosphere may lead to unique observable EM signatures of these transients in AGN disks. SN and kilonova emissions in AGN disks could be outshone by the AGN disk emissions \citep{zhu2021thermonuclear,zhu2021neutron}, while shock breakout signals and interaction emissions between ejecta and disk atmosphere can be bright enough to be detected \citep{zhu2021thermonuclear,grishin2021supernova,ren2022}. The EM signals of AGN GRBs are highly diverse, being dependent on the burst properties, the disk structure, and the location of bursts \citep{cheng1999,perna2021,zhu2021neutron,kimura2021,lazzati2022,ray2022,wang2022,yuan2022}. Although binary BH (BBH) mergers are not expected to directly generate EM emission\footnote{\cite{zhang2016} predicted that mergers of charged BHs could make EM counterparts, e.g., fast radio bursts or short GRBs.}, BBH mergers embedded in AGN disks can potentially power EM counterparts through accretion and interaction with disk gas \citep{bartos2017,mckernan2019,wang2021accretion1,kimura2021,tagawa2022}. Interestingly, \cite{graham2020} reported a rapidly evolving AGN transient candidate, which was plausibly associated with a BBH event detected by the LIGO-Virgo Collaboration \citep[i.e., GW190521;][]{LIGO190521}.

Besides observable EM signals, AGN disk transients are expected to be important sources of high-energy astrophysical neutrinos. \cite{zhu2021high2} studied the evolution of GRB jets embedded in AGN disks and found that detectable TeV--PeV neutrinos would be produced efficiently through photomeson interactions when GRB jets are choked by the dense disk atmosphere. By analogy with SNe Ia of which the ejecta can have an interaction with the dense circum-stellar medium to generate neutrino production \citep{waxman2001tev,waxman2016shock,murase2011new,murase2018new,murase2019high,li2019pev,wang2019transient}, \cite{zhu2021high} predicted that the interaction between disk atmosphere and ejecta from AGN disk transients would lead to high-energy neutrino emissions via  inelastic $pp$ interactions. Since stable gas-dominated accretion disks have a Gaussian density distribution, much sharper compared with that of the circum-stellar medium around SN II,  \cite{zhu2021high} showed that the neutrino emission from AGN disk transients would be burst-like, with predicted duration much shorter than that of SNe II \citep{murase2018new}. As suggested by the X-ray spectra of AGNs that usually show the imprint of absorption from ionized gas, circum-disk media, including AGN outflows or disk winds, are believed to be ubiquitous around the AGN disks \citep{1984ApJ...281...90H,1995MNRAS.273.1167R}. However, a detailed model for the effect of circum-disk media on EM and neutrino emissions from AGN disk transients is still lacking.

Outflows have been found in $\gtrsim50\%$ of nearby AGNs, by which the central SMBH will be connected with its host galaxy \citep{2015ARA&A..53..115K,2021NatAs...5...13L}. Powerful outflows carrying a significant fraction of AGN power can affect the evolution of their host galaxies through `negative' feedback \citep{2021NatAs...5...13L}. Based on the blueshifted X-ray absorption lines, outflow velocities could range from several hundred kilometers per second to ultrafast speeds $\sim 0.1-0.25\,c$ \citep{2003MNRAS.345..705P,2005A&A...431..111B,2007MNRAS.379.1359M,2021MNRAS.503.1442M}. These outflows or disk winds may influence the gas density distribution in the vicinity of the AGN disk \citep{proga2000dynamics,proga2004dynamics,yoshioka2022large}. Consequently, for a transient explosion inside AGN disks, the dynamics of ejecta would be different and the production of high-energy neutrinos would be changed. %Due to the existence of AGN outflow or disk wind, we consider various density profiles to explore their influence on neutrino production. We find that the neutrino production will be enhanced significantly and make a much higher contribution to the diffuse neutrino background thanks to the extra circum-disk materials.
In this paper, we take the first steps in modeling neutrino production from AGN disk transients by considering different circum-disk environments.

The paper is organized as follows. We introduce the structure of AGN disks and circum-disk medium in Section~\ref{sec:struc}. Then we calculate the shock dynamics and the lightcurves of neutrino emissions in Section~\ref{sec:NP}. In Section~\ref{sec:fluence}, the neutrino emission fluences and the contribution to diffuse neutrino background are presented. Lastly, we provide a discussion in Section~\ref{sec:discusssion}.

\section{Structure of AGN Disk and Circum-disk Medium}\label{sec:struc}

We use radial distance $r$ and vertical height $h$ relative to the mid-plane to describe the structure of AGN disks and circum-disk gas, where $r$ is expressed in units of the SMBH's Schwarzschild radius $r_{\rm{S}}=2GM_\bullet/c^2$ with the gravitational constant $G$, the mass of SMBH $M_\bullet$, and the speed of light $c$. The AGN disk atmosphere is assumed to be a stable gas-dominated disk whose vertical density distribution obeys a Gaussian density profile \citep[e.g.,][]{netzer2013physics}. The gas materials in the vicinity of the disk can have an extended distribution due to the existence of mass outflows or disk winds around the AGNs \citep[see e.g.,][for simulations]{proga2000dynamics,proga2004dynamics}.

\subsection{spherically symmetric wind}

Firstly, we assume the circum-disk materials are stationary and spherically symmetric radiative line-driven winds. Then, the outflow mass rate can be estimated for a uniform radial outflow of velocity $v_w$, i.e.,

\begin{equation}
\dot M = 4\pi {r^2}b\rho_{\rm w} (r)v_{\rm w},
\end{equation}
where $b$ is the covering factor of the wind and $\rho_{\rm w} (r)$ is the gas density at a radial distance $r$. The covering factor $b$ is generally large or even close to unity, e.g., $b\simeq 0.75 \pm 0.25$ for PG1211+143 \citep{2007MNRAS.374..823P}. Therefore, the gas density can be evaluated by

\begin{equation}
    \rho_{\rm{w}}(r) \simeq10^{-15}b_0^{-1}\dot{M}_{25}v_{\rm w,8}^{-1}(r/10^3r_{\rm {S}})^{-2}\,{\rm g}\,{\rm cm}^{-3}
    \label{eqrho}
\end{equation}
with a mass-loss rate of $\dot{M}_{25}\sim10^{25}\,{\rm g}\,{\rm s}^{-1}\sim0.36\,M_\odot\,{\rm yr}^{-1}$ and a wind velocity of $v_{\rm w,8}=10^{8}\,{\rm cm}\,{\rm s}^{-1}$. Hereafter, the convention $Q_x=Q/10^x$ is adopted in cgs units. The mass-loss rate is comparable with the Eddington accretion rate $\dot M_\mathrm{Edd} \simeq 0.33 \, M_{\sun}\,\rm yr^{-1}$ for an SMBH mass of $M_\bullet=10^{7}M_\sun$ accreting at an efficiency of $10\%$,  whereas the corresponding mechanical energy is much lower than the Eddington luminosity with a ratio $\sim 4\times 10^{-4}$ consistent with the prediction of continuum driving \citep[see e.g.,][]{2015ARA&A..53..115K}. 

For a stationary and spherically symmetric wind, we thus empirically define the combined vertical density profile of disk atmosphere and circum-disk material as a function of $r$ and $h$, i.e., 

\begin{equation}
	\begin{aligned}
		\rho_{\rm{d}}(r,h)=\begin{cases}
			\rho_0(r)\exp(-h^2/2H^2), & h<h_{\rm{c}},\\
			\rho_0(r)\exp(-h_{\rm{c}}^{2}/2H^2)\left( \frac{r'}{r'_{\rm{c}}} \right)^   {-2},& h>h_{\rm{c}},\\
		\end{cases}
	\end{aligned}\label{eqdist1}
\end{equation}
where $h_{\rm{c}}$ is the critical height boundary at which the gas density of the disk atmosphere is equal to that of circum-disk materials. Beyond $h_{\rm{c}}$, the gas density becomes dominant by the circum-disk materials rather than the disk atmosphere with a Gaussian distribution. $r'$ and $r'_c$ indicate the radial distance from the SMBH for the vertical height $h$ and the critical height $h_{\rm{c}}$, i.e., $r' = \sqrt {{r^2} + {h^2}} $ and ${r'_c} = \sqrt {{r^2} + h_{\rm c}^2}$. The mid-plane radial density $\rho_0(r)\approx1.24\times10^{-7}\,M_{\bullet,7}^{-2}(r/10^3\,r_{\rm{S}})^{-3}\,{\rm{g}}\,{\rm{cm}}^{-3}$ and disk scale height compared to the radial distance $H/r\approx8\times10^{-3}(r/10^3\,r_{\rm{S}})^{1/2}$ are adopted from the disk model suggested by \cite{sirko2003spectral}, where $M_{\bullet,7}=M_\bullet/10^7\,M_\odot$. The expressions of $\rho_0(r)$ and $H/r$ are valid for the radial region of $10^3 \lesssim  r/r_{\rm S}\ \lesssim 10^5$ that we are interested in our study. For a specific AGN disk transient occurring at the mid-plane of the disk with a specific radius $r$ from the SMBH with the mass of $M_\bullet$ (i.e., the fixed $\rho_{0}(r)$ and $H$), we can introduce a critical density $\rho_{\rm{c}}=\rho_{0}(r)\exp(-h^2_{\rm{c}}/2H^2)$ to conveniently judge the dominance of the disk atmosphere or the circum-disk materials during the shock dynamical evolution. Therefore, we treat $\rho_{\rm{c}}$ as a free parameter instead of $h_{\rm c}$.

\subsection{cylindrically symmetric wind}
Although for the larger region around the SMBH, the AGN disk wind tends to be quasi-spherical from the SMBH, in the vicinity of the AGN disk, the disk wind streamlines will be more nearly vertical for the lower velocities of disk winds \citep{1995ApJ...451..498M,1997ApJ...474...91M}. In addition, the vertically launched wind can maintain its direction if the radiation force is effective enough, even though it may bend toward a radial direction at a large height \citep{2010A&A...516A..89R}. The angle of the wind streamlines with the mid-plane of the disk can be large so that the disk wind moves approximately vertically if $h<r$, eventually forming an approximately cylindrically symmetric circum-disk materials distribution in the vicinity of the AGN disk. In order to more comprehensively explore the influence of circum-disk materials on neutrino production, in this situation, for the extended region, the gas distribution is assumed to have a power-law density profile in the vertical direction. We empirically define the combined vertical density profile of disk atmosphere and circum-disk material as 

\begin{equation}
	\begin{aligned}
		\rho_{\rm{d}}(r,h)=\begin{cases}
			\rho_0(r)\exp(-h^2/2H^2), & h<h_{\rm{c}},\\
			\rho_0(r)\exp(-h_{\rm{c}}^{2}/2H^2)\left( \frac{h}{h_{\rm{c}}} \right)^   {-\alpha},& h>h_{\rm{c}},\\
		\end{cases}
	\end{aligned}\label{eqdist2}
\end{equation}
where $\alpha$ is the power-law index of circum-disk gas's density. The density in the wind scales as $\rho  \propto 1/{v_{\rm w}(h)}$ in the vertical direction. The disk wind will be accelerated by the radiation force and the acceleration can be fast and continue until $\sim 500-1000\,r_{\rm S}$ \citep{2010A&A...516A..89R}. Although the realistic acceleration is complicated, we adopt the power-law index $\alpha=[1.5,2,3]$ to explore this influence on neutrino production. When initially launched, The wind gas density at $r=10^3\, r_{\rm S}$ will be comparable with Equation (\ref{eqrho}) for the same accretion rate, the SMBH mass, and the wind velocity, i.e., $\rho_{\rm w} \sim 10^{-15}\,\rm g\,cm^{-3}$ for $v_{\rm w}=10^{8}\,{\rm cm}\,{\rm s}^{-1}$. Besides, we also consider smaller wind velocities that can produce nearly vertical wind streamlines easier \citep{1995ApJ...451..498M,1997ApJ...474...91M}, say, $\rho_{\rm w} \sim 10^{-14}\,\rm g\,cm^{-3}$ for $v_{\rm w}=10^{7}\,{\rm cm}\,{\rm s}^{-1}$ and $\rho_{\rm w} \sim 10^{-13}\,\rm g\,cm^{-3}$ for $v_{\rm w}=10^{6}\,{\rm cm}\,{\rm s}^{-1}$ near the wind initially launching position. Considering the diverse wind velocities, we adopt the critical density $\rho_{\rm c}=[10^{-13},10^{-14},10^{-15}]\,\rm g\,cm^{-3}$ to explore their influence on neutrino production instead of $h_{\rm c}$ as same as for spherically symmetric wind. As presented in Table~\ref{Tab:1}, $\rho_{\rm c}=10^{-13}\,\rm g\,cm^{-3}$ is large enough to consume almost all shock energy and thus a larger $\rho_{\rm c}$ will not enhance the neutrino production significantly (please see the details in next two Section). Moreover, $\rho_{\rm c}=10^{-15}\,\rm g\,cm^{-3}$ is small enough so that the neutrino fluence tends to be the same as the pure Gaussian density profile as shown in Figure~\ref{fig:fluence}. As a result, we do not employ the $\rho_{\rm c}$ larger than $10^{-13}\,\rm g\,cm^{-3}$ or smaller than $10^{-15}\,\rm g\,cm^{-3}$.

Then we explore the shock dynamics and the neutrino production under the gas density distribution described as in Equations (\ref{eqdist1}) and (\ref{eqdist2}) when an AGN disk transient occurs and the line-of-sight is perpendicular to the disk. The outermost height of circum-disk materials is limited within $10^3 \,r_\mathrm{S}$ as suggested by the numerical simulations \citep[e.g.,][]{proga2000dynamics}, i.e., $h \le 10^3\,r_\mathrm{S}$. Besides, we neglect the velocities of circum-disk materials during the interaction between the ejecta of a transient source and the circum-disk gas since the velocity of ejecta is generally much larger than the velocity of circum-disk gas for the adopted parameter values here.

\section{neutrino production}\label{sec:NP}
\subsection{Shock Dynamics}

When a transient explosion occurs embedded in an AGN disk with energy $E_0$ and ejecta mass $M_{\rm ej}$, a forward shock and a reverse shock at the interface between ejecta and gas can be produced as the ejecta crashes into the disk atmosphere and circum-disk gas. Because the high-energy neutrino production from the reverse shock is much weaker \citep[e.g.,][]{murase2011new}, we focus on the neutrino contributions from the forward shock. To simplify, all AGN disk transients are considered to be exploded at the mid-plane of disks. Thus, for a transient located at a radial distance $r$ away from the central SMBH, the ejecta shock velocity can be given by \citep{matzner1999expulsion}

\begin{equation}
v_{\rm s}\left( h \right) \approx \left( \frac{E_0}{M_{\mathrm{ej}}+M_{\rm sw}\left( h \right)} \right) ^{1/2}\left( \frac{\rho _{\mathrm{d}}\left( h \right)}{\rho _0} \right) ^{-\mu},
\end{equation}
where $\mu=0.19$ and $M_{\rm sw}\left( h \right) \approx 4\pi \int_0^h{\rho \left( h' \right) h'^2dh'} $ is the swept mass. %In this article we use a fixed upper numerical simulation limit of $h$ is $1000r_{\rm S}$. The kinetic luminosity of the shock at $h$ can be calculated by $L_{\rm s}=2\pi \rho _dv_{s}^{3}h^2$ and the time when shock move to $h$ is $t\left( h \right) \approx \int_0^h{\frac{dh}{v_{\rm s}\left( h \right)}}$. 
The time when the shock moves to $h$ can be calculated by $t(h)\approx\int^{h}_0dh'/v_{\rm{s}}(h')$ and, hence, $v_{\rm s}$ can be also expressed as a function of time $t$. With known shock velocity, we can get the shock kinetic luminosity 

\begin{equation}\label{kinetic}
    L_{\rm s} \approx 2\pi\rho_{\rm d}v_{\rm s}^3h^2.
\end{equation}

The shock would be radiation-mediated when the optical depth of Thomson scattering $\tau>c/v_s$ so that the particle acceleration is prohibited \citep{waxman2001tev,murase2011new,katz2011x}. Therefore, particle acceleration and neutrino production can not occur prior to the shock breakout ($\tau \approx c/v_s$). When photons start to escape after shock breakout at $h_{\rm bo}$, the ejecta shock converts from a radiation-dominated shock into a collisionless shock. The shock breakout height $h_{\rm bo}$ can be solved by ${{c}/{v_{\rm s}\left( h_{\rm bo} \right)}}\approx \tau \left( h_{\rm bo} \right) \approx\int_{h_{\rm bo}}^{+\infty}{\kappa \rho \left( h \right) dh}$, where a constant electron scattering opacity of solar composition for $\kappa \approx0.34\,\mathrm{cm}^2\mathrm{g}^{-1}$ is adopted. The maximum shock velocity is set to $v_{\rm s,max}\approx2.1\,v_{\rm bo}$ based on \cite{matzner1999expulsion}.

\subsection{Maximum Proton Energy}\label{secmpe}

Particle acceleration and neutrino production can occur when the shock becomes a collisionless shock after the breakout. The accelerated proton adopted is a non-thermal power-law distribution with a high energy exponential cutoff, i.e., $d{n_p}/d{\epsilon _p} \propto \epsilon _p^{ - s}\exp ( - {\epsilon _p}/{\epsilon _{p,\max }})$ with $s \approx 2$, where $\epsilon_p$ and $\epsilon_{p,{\rm max}}$ are the proton energy and maximum proton energy, respectively. The proton acceleration timescale is $t_{{\rm acc}}=\eta \epsilon_p/eBc$, where $\eta\sim20c^2/3v_{\rm s}^2$ is for the Bohm diffusion \citep{1983RPPh...46..973D,murase2018new} and $e$ is the electron charge. $B=\sqrt{4\pi\varepsilon_B\rho_{\rm d}v_{\rm s}^2}$ is the magnetic ﬁeld strength in the shocked disk and circum-disk materials, where a typical value of the magnetic ﬁeld energy fraction is used $\varepsilon_B = 0.01$. Then, high-energy neutrinos can be produced by the $pp$ collisions. Only the $pp$ interaction and the adiabatic cooling can exert a prominent impact on the maximum proton energy, and they constrain the maximum proton energy at different energy ranges. By only considering the $pp$ interaction, the cooling timescale is $t_{pp}=1/[(\rho _{\rm s}/m_p)\kappa_{pp}\sigma _{pp}c]$, where $\rho_{\rm s}=4\rho_{\rm d}$ is the density of the shocked materials, $\kappa_{pp}\approx0.5$ is the $pp$ inelasticity, and $\sigma_{pp}\approx5\times10^{-26}\,\rm{cm^{-2}}$ is the $pp$ cross section \citep{ParticleDataGroup:2020ssz}. The maximum energy of protons by comparing the acceleration timescale and $pp$ cooling timescale is $\epsilon _{p,\max}^{pp}\approx 9.4\times 10^3 \rho _{\rm d,-15}^{-1/2}v_{\rm s,9}^{3}\,\mathrm{TeV}$. If the proton acceleration timescale is limited by the adiabatic cooling timescale, i.e., $t_{pp} = t_{\rm ad} \approx h/v_{\rm s}$, the proton maximum energy is $\epsilon _{p,\max}^{\rm ad}\approx 1.7\times 10^3\,\rho _{\rm d,-15}^{1/2}v_{\rm s,9}^{2}h_{14}\,\mathrm{TeV}$. Thus, we can finally get the maximum proton energy as $\epsilon _{p,\max}\approx \min \left( \epsilon _{p,\max}^{pp},\epsilon _{p,\max}^{\rm ad} \right)$.

\subsection{Neutrino Emission Lightcurves} \label{sec:NeutrinoProduction}

The neutrino luminosity via the $pp$ interaction can be expressed as

\begin{equation}
	\epsilon _{\nu}L_{\nu}\approx \frac{3A}{4\left( 1+A \right)}\frac{\min \left( 1,f_{pp}\right) \varepsilon _{cr}L_{\rm s}}{\ln \left( \epsilon _{p,\max}/\epsilon _{p,\min} \right)},
\end{equation}
where $\epsilon_\nu\approx 0.05\,\epsilon _p$ is the neutrino energy, $\varepsilon _{\rm cr}=0.1$ is the fraction of energy carried by cosmic rays \citep{caprioli2014simulations}, $A=2$ denotes the average ratio of charged to neutral pions for the $pp$ interaction, $f_{pp}\approx t_{\rm ad}/t_{pp}\approx0.18\,\rho _{\rm d,-15}v_{\rm s,9}^{-1}h_{14}$ is the estimated efficiency of the $pp$ interaction \citep[e.g.,][]{zhu2021high}, and $\ln \left( \epsilon _{p,\max}/\epsilon _{p,\min} \right)$ is the normalization factor with the minimum proton energy $\epsilon _{p,\min}\approx m_pc^2$ and the proton mass $m_p$.

\begin{figure}
	\centering
	\includegraphics[width = 0.99\linewidth , trim = 25 5 55 30,clip]{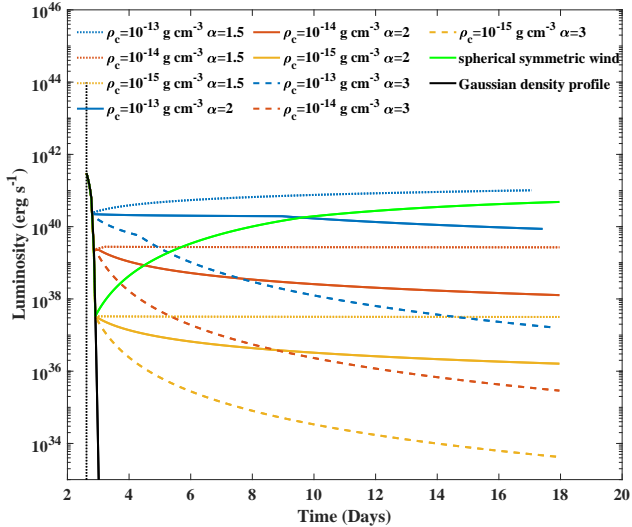}
	\caption{Lightcurves of neutrino emission at $\epsilon _{\nu}=1\,\mathrm{TeV}$ for an AGN SN Ia occurring at $r=10^3\,r_{\rm S}$ away from a $10^7\,M_{\odot}$ SMBH. The vertical black dotted line indicates the shock breakout time after the stellar explosion. Blue, red and yellow lines represent different critical densities $\rho_{\rm c} = 10^{-13}$, $10^{-14}$, and $10^{-15}$, respectively. The solid, dashed and dotted lines display different $\alpha$ of $2,$ $3,$ $1.5$, respectively. The green line is corresponding to the spherically symmetric wind described by Equation~(\ref{eqdist1}) with $\rho_{\rm c}=10^{-15}\,\rm g\, cm^{-3}$. The colored lines are covered by the black solid line before they deviate from the pure Gaussian density profile. }
	\label{fig:Lightcurves}
\end{figure}

\begin{figure}
	\centering
	\includegraphics[width = 0.99\linewidth , trim = 15 5 55 30, clip]{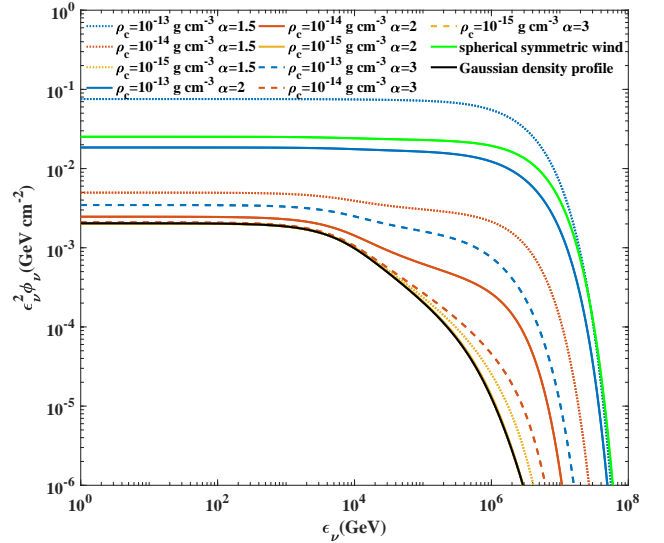}
	\caption{Energy fluences of all-flavor neutrinos for a single AGN SN Ia event occurring at $D_{\rm L} = 10 \,\rm Mpc$. The line styles are the same as those in Figure~\ref{fig:Lightcurves}. Note that the yellow solid and dashed lines are covered by the black solid line due to the negligible circum-disk media and become invisible in the figure.}
	\label{fig:fluence}
\end{figure}

We adopt an SN Ia with typical energy $E_0=10^{51}$\,erg and ejecta mass $M_{\rm ej}=1.3\,M_{\odot}$ occurring at $r=10^3\,r_{\rm S}$ around a SMBH of mass $M_{\bullet}=10^7M_{\odot}$ as the fiducial model hereafter. We present lightcurves of neutrino emission from a single source at 1\,TeV in Figure \ref{fig:Lightcurves} with different settings of $\alpha=[1.5,2,3]$ and $\rho_{\rm c}=[10^{-13},10^{-14},10^{-15}]\,\rm{g\,cm^{-3} }$ for the cylindrically symmetric circum-disk materials distribution and $\rho_{\rm c}=10^{-15}\,\rm g\, cm^{-3}$ for the spherically symmetric distribution. The breakout time for each group of parameters is almost the same $\simeq 2.6$\,days after the AGN disk transients since the shock breakout height is still dominated by the AGN disk materials rather than the circum-disk materials. In addition, for the same critical height $h_{\rm c}$ (or $\rho_{\rm c}$), before $h_{\rm c}$, the neutrino emission lightcurves are determined by the AGN disk and therefore are initially almost identical. The tendencies of lightcurves before the critical height $h_{\rm c}$ are similar to the pure Gaussian density profile. Since the $pp$ interaction efficiency is in direct proportion to the gas density, the neutrino emission luminosity decreases sharply before $h_{\rm c}$, consistent with the expectation of rapid descent for the Gaussian density distribution. Beyond $h_{\rm c}$, the circum-disk materials start to dominate. The gas density drops off more gently compared with the single Gaussian density distribution as the shock height, inducing the slower descent of the neutrino emission or even rise of neutrino luminosity as mainly determined by the shock kinetic luminosity with Equation~(\ref{kinetic}). For the circum-disk materials described by Equation~(\ref{eqdist2}), the neutrino emission decreases more steeply for a larger index $\alpha$. For the spherically symmetric circum-disk materials, the neutrino luminosity is higher than in the cylindrically symmetric case with the same critical density $\rho_{\rm c}$ since its gas density decreases more slowly in the vertical direction.

\begin{deluxetable*}{ccc|cccc|cccc|cccc}[htpb!] \label{Tab:1}
	\tablecaption{Parameters for AGN Disk Transients}
	\tablecolumns{15}
	\tablewidth{0pt}
	
	\setlength{\tabcolsep}{0.8mm}{}
	
	\tablehead{\colhead{Transients} &\colhead{$E_{0,51}$} &\colhead{$M_{\rm ej}$} &\colhead{$M_{\bullet}$} &\colhead{$r/r_{\rm S}$} &\colhead{$\rho_{c,-15}$} & \colhead{$\alpha$} &\colhead{$\rho_{\rm bo,-11}$} &\colhead{$h_{\rm bo,14}$} &\colhead{$E_{S,48}$} &\colhead{$\epsilon_{p,\max}^{\rm bo}/\mathrm{TeV}$} &\colhead{$\underset{(-5^{\circ}<\delta<30^{\circ})}{N_{\nu _{\mu}}}$ }
    &\colhead{$\underset{(30^{\circ}<\delta<90^{\circ})}{N_{\nu _{\mu}}}$ }
    &\colhead{$\underset{(-30^{\circ}<\delta<-5^{\circ})}{N_{\nu _{\mu}}}$ }
    &\colhead{$\underset{(-90^{\circ}<\delta<-30^{\circ})}{N_{\nu _{\mu}}}$} }
	\startdata
SN Ia & 1 & 1.3 & $10^7$ & $10^3$ & 100 & 1.5 & 1.2 & 1 & 522 & 113 & 1.4498 & 1.0454 & 0.1846 & 0.0397\\
SN Ia & 1 & 1.3 & $10^7$ & $10^3$ & 10 & 1.5 & 1.6 & 1 & 63 & 80 & 0.0704 & 0.0538 & 0.0075 & 0.0015\\
SN Ia & 1 & 1.3 & $10^7$ & $10^3$ & 1 & 1.5 & 1.6 & 1 & 15 & 80 & 0.0142 & 0.0132 & 0.0005 & 0.00004\\
SN Ia & 1 & 1.3 & $10^7$ & $10^3$ & 100 & 2 & 1.4 & 1 & 154 & 98 & 0.3364 & 0.2449 & 0.0418 & 0.0089\\
SN Ia & 1 & 1.3 & $10^7$ & $10^3$ & 10 & 2 & 1.6 & 1 & 26 & 80 & 0.0218 & 0.0188 & 0.0014 & 0.0002\\
SN Ia & 1 & 1.3 & $10^7$ & $10^3$ & 1 & 2 & 1.6 & 1 & 12 & 80 & 0.0136 & 0.0128 & 0.0005 & 0.00003\\
SN Ia & 1 & 1.3 & $10^7$ & $10^3$ & 100 & 3 & 1.5 & 1 & 28 & 85 & 0.0412 & 0.0332 & 0.0034 & 0.0006\\
SN Ia & 1 & 1.3 & $10^7$ & $10^3$ & 10 & 3 & 1.6 & 1 & 12 & 80 & 0.0148 & 0.0136 & 0.0006 & 0.00005\\
SN Ia & 1 & 1.3 & $10^7$ & $10^3$ & 1 & 3 & 1.6 & 1 & 10 & 80 & 0.0136 & 0.0127 & 0.0005 & 0.00003\\
\hline
%要不要换几个探测数大的参数
SN Ia & 1 & 1.3 & $10^6$ & $10^3$ & 10 & 2 & 4.19 & 0.12 & 14 & 5167 & 0.0017 & 0.0013 & 0.0002 & 0.00003\\
SN Ia & 1 & 1.3 & $10^8$ & $10^3$ & 10 & 2 & 0.78 & 7.52 & 389 & 0.47 & 0.5790 & 0.5164 & 0.0228 & 0.0010\\
\hline
SN Ia (SS) & 1 & 1.3 & $10^7$ & $10^3$ & 1 & -- & 1.6 & 1 & 180 & 80 & 0.4744 & 0.3397 & 0.0636 & 0.0143\\
\hline
Kilonova & 1 & 0.05 & $10^7$ & $10^3$ & 10 & 2 & 1.54 & 1.0 & 29 & 98 & 0.0241 & 0.0207 & 0.0016 & 0.0002\\
CCSN & 3 & 10 & $10^7$ & $10^3$ & 10 & 2 & 1.11 & 1.0 & 49.8 & 305 & 0.0410 & 0.0339 & 0.0030 & 0.0005\\
GRB-SN & 30 & 10 & $10^7$ & $10^3$ & 10 & 2 & 0.27 & 1.09 & 670 & 44392 & 0.2186 & 0.1529 & 0.0327 & 0.0079\\
	\enddata
	\tablecomments{The columns are [1] the kind of AGN disk transients; [2] explosion energy (in $10^{51} \,\rm erg \,s^{-1}$); [3] ejecta mass (in $M_\odot$); [4] SMBH mass (in $M_\odot$); [5] radial location; [6] disk density profile breaking point; [7] density attenuation index above the broken height; [8] disk density at the shock breakout (in $10^{-11}\,\rm g\,cm^{-3}$); [9] breakout vertical height (in $10^{14}\rm cm$); [10] shock energy after the shock breakout; [11] the maximum proton energy when the shock breaks out (in unit of TeV); [12-15] the expected (anti-)muon neutrino event number (in 100\,GeV--100\,PeV) by IceCube from four groups of incidence declination angles for a single event at $D_{\rm L} = 10 \,\rm Mpc$, and the IceCube effective areas for diverse declination angles are adopted as provided in \citep{aartsen2020time}. SN Ia (SS) means that the SN Ia ejecta propagates inside the spherically symmetric circum-disk materials. }
\end{deluxetable*}

In addition, we explore the impact of diverse AGN accretion disks (the SMBH mass) and diverse AGN disk transients on the neutrino emission and the detectability as listed in Table~\ref{Tab:1}.

\section{Neutrino Fluence and Diffuse Neutrino Emission}\label{sec:fluence}

The all-flavor neutrino fluence for a single event can be written as

\begin{equation}
	\epsilon _{\nu}^{2}\phi _{\nu}\approx \frac{1}{4\pi D_{L}^{2}}\int_{t_{\rm bo}}^{+\infty}{\epsilon _{\nu}L_{\nu}dt},
\end{equation}
where $dt=dh/v_{\rm s}$, $D_{\rm L}$ is the luminosity distance, and $t_{\rm bo}$ is the shock breakout time. We show the neutrino fluence with model parameters $M_{\bullet}=10^7M_{\odot}$ and $r=10^3r_{\rm S}$ in Figure \ref{fig:fluence}.

For each case, we estimate the number of (anti-)moun neutrino $\nu_{\mu}+ {\bar\nu_{\mu}}$ events detected by IceCube as 
\begin{equation}
	%N_{\nu _{\mu}}\left \right) =\int_{1\mathrm{TeV}}^{\epsilon _{\nu %,\max}}{d\epsilon _{\nu _{\mu}}A_{\mathrm{eff}}\left( \epsilon _{\nu _{\mu}} \right) %\phi _{\nu _{\mu}}},
    N_{\nu _{\mu}}=\int{d\epsilon _{\nu _{\mu}}A_{\mathrm{eff}}\left( \epsilon _{\nu _{\mu}} \right) \phi _{\nu _{\mu}}},
	\label{eq5}
\end{equation}
where the (anti-)muon neutrino fluence is one-third of the all-flavor neutrino fluence considering the neutrino oscillation, i.e., $\phi _{\nu _{\mu}}=\frac{1}{3}\phi _{\nu}$, and $A_{\mathrm{eff}}\left( \epsilon _{\nu _{\mu}} \right)$ is the effective area (100\,GeV--100\,PeV) for a point source given by \citep{aartsen2020time}\footnote{The adopted IceCube effective area is based on Figure 1 in the Supplemental Material of \citep{aartsen2020time} and we obtain four groups of effective areas in the range of 100\,GeV--100\,PeV for diverse declination angles through the data point capture along the lines.}. %\citep{aartsen2017extending}.
We calculated the total shock energy after the shock breakout (i.e., $E_{\rm s}=\int_{t_{\rm bo}}^{+\infty}{L_{\rm s}dt}$ ) for  reference. For the circum-disk medium distribution with $\rho_{\rm c}=10^{-13}\,\mathrm{g}\, \mathrm{cm}^{-3}$ and $\alpha=2$, the neutrino fluence below 10\,TeV is about one order of magnitude larger than the neutrino emission in Gaussian density profile accretion disk, and for the optimistic case with $\rho_{\rm c}=10^{-13}\,\mathrm{g}\, \mathrm{cm}^{-3}$ and $\alpha=1.5$, the neutrino emission can be around forty times larger than the pure Gaussian density profile accretion disk.
%The fluence for $\rho _c=10^{-14}\,\,\mathrm{g} \mathrm{cm}^{-3}$ and $\alpha=1.5$ is basically identical to $\rho _c=10^{-13}\,\,\mathrm{g} \mathrm{cm}^{-3}$ and $\alpha=2$. 
Basically, a higher neutrino fluence can be expected for a denser gas environment due to the higher efficiencies of $pp$ collisions in Figure~\ref{fig:fluence}. As shown in Figure~\ref{fig:fluence}, for the relatively rarefied circum-disk media, e.g., $\rho_{\rm c}=10^{-15}\,\mathrm{g}\, \mathrm{cm}^{-3}$ with $\alpha=2$ (yellow solid) or $\alpha=3$ (yellow dashed), the effect of circum-disk media is so weak that the neutrino fluence tends to be the same as the case of pure Gaussian density profile. At larger height $h$, the maximum proton energy is generally determined by the adiabatic cooling of the shock. For a lower gas density, e.g., larger $\alpha$ for the same $\rho_{\rm c}$ or smaller $\rho_{\rm c}$ for the same $\alpha$, the maximum proton energy tends to be lower, inducing a smaller neutrino cutoff energy. The expected number of (anti-)muon neutrinos from a single source with different $\alpha$ and $\rho_{\rm c}$ has been calculated and listed in Table \ref{Tab:1}. The results are basically consistent with the prediction that higher $\rho_{\rm c}$ and  lower $\alpha$ lead to a higher neutrino detection event rate.

We explore the different kinds of AGN disk transients including SN Ia, CCSN, GRB-SN, and kilonova with the consideration of their classical explosion energies and ejecta masses. A large fraction of stars in the AGN disk are formed in the self-gravitating region, i.e., $10^3 \lesssim r/r_{\rm S} \lesssim 10^5$ \citep{sirko2003spectral,thompson2005radiation}. AGN stars will experience radial migration from outer to inner of the disks, whereas most of them may not be able to migrate to the inner trapped orbits of the disk ($\lesssim 10^3\,r_{\rm S}$) before their deaths or within the AGN lifetime. As a result, we mainly consider these AGN disk transients occurring at the self-gravitating region of the disk. SMBHs usually have a mass of  $\sim10^6-10^8M_{\odot}$ with a nearly uniform local mass function \citep{2014Mass}. Here we simply set all transients occurring around an SMBH with a mass of $10^7\,M_{\odot}$ and at the typical radial locations of $r = 10^3\,r_{\rm S}$. From Table \ref{Tab:1}, we can see that a transient with a higher explosion energy, e.g., CCSN and GRB-SN, can drive a more powerful shock and consequently produce a higher neutrino fluence and a larger maximum neutrino energy. At $D_{\rm L}=10$\,Mpc, the expected (anti-)muon neutrino events from these AGN disk transients are shown in Table~\ref{Tab:1} as well.

Note that IceCube developed its low-energy infill array DeepCore extending the neutrino detection energy to $\sim$10\,GeV \citep{2017EPJC...77..146A}. The effective area of the DeepCore sub-array is roughly obeying $A_{DC} \propto {\epsilon_\nu}^{3}$ \citep{2017EPJC...77..146A}. Since our obtained neutrino fluence is flat at the lower energy range, the neutrino number per square centimeter is approximately following $\epsilon_\nu \phi _{\nu _{\mu}} \propto {\epsilon_\nu}^{-1}$. Therefore, the expected neutrino number by the DeepCore sub-array is proportion to ${\epsilon_\nu}^{2}$ so that its contribution (mainly at $<100\,\rm GeV$) to the total expected neutrino number is less even than IceCube main array at 100\,GeV for a flat neutrino flux. As a result, we focus on the neutrino with energies of 100\,GeV--100\,PeV detected by the IceCube main array.

\begin{figure}
	\centering
	\includegraphics[width = 0.99\linewidth , trim = 15 5 55 30, clip]{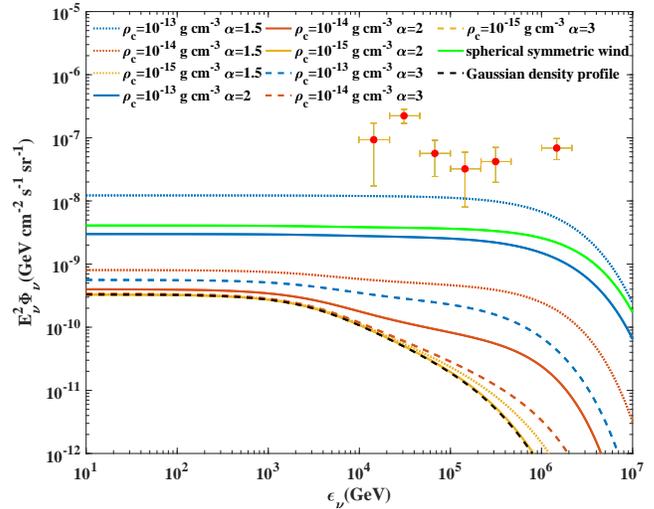}
	\caption{The upper limits of expected all-flavor diffuse neutrino fluences contributed from AGN SN Ia. The red points are observed astrophysical diﬀuse neutrino background by IceCube \citep{2015ApJ...809...98A}.}
	\label{fig:diffuse}
\end{figure}

The diffuse neutrino emission can be calculated by

\begin{equation}
	E_{\nu}^{2}\Phi _{\nu}=\frac{c}{4\pi H_0}\int_0^{z_{\max}}{\frac{R_0S\left( z \right) \mathcal{E} \left( \epsilon _{\nu} \right)}{\left( 1+z \right) ^2\sqrt{\Omega _{\rm m}\left( 1+z \right) ^3+\Omega _{\Lambda}}}},
\end{equation}
where $z$ is the redshift, $R_0$ is the theoretical local event rate density of AGN disk transient which is listed in Table~\ref{Tab:2}, $\mathcal{E} \left( \epsilon _{\nu} \right) \approx \int_{t_{\rm bo}}^{+\infty}{\epsilon _{\nu}L_{\nu}dt}$ is the local neutrino energy budget for a single explosion, $E_{\nu}=\epsilon _{\nu}/\left( 1+z \right)$ is the neutrino energy in the observer’s frame. The standard $\Lambda \mathrm{CDM}$ cosmology with $H_0=67.8\, \mathrm{km} \mathrm{s}^{-1}\,\,\mathrm{Mpc}^{-1}$, $\Omega _{\rm m}=0.308$ and $\Omega _{\Lambda}=0.692$ \citep{2016A&A...594A..13P} is applied. We assume that the source evolution traces the cosmological star formation history,  which can be expressed by a redshift-evolution factor as $S\left( z \right) =\left[ \left( 1+z \right) ^{3.4\eta}+\left( \frac{1+z}{5000} \right) ^{-0.3\eta}+\left( \frac{1+z}{9} \right) ^{-3.5\eta} \right] ^{1/\eta}$ \citep{sun2015extragalactic}, where $\eta =-10$ \citep{yuksel2008revealing}.

The predicted diffuse neutrino ﬂuences of AGN SN Ia are shown in Figure \ref{fig:diffuse}.  We calculate the contribution of SN Ia, kilonova, CCSN, and GRB-SN to the diffuse neutrino background given by IceCube. We present the results of SN Ia and other AGN disk transients in Table~\ref{Tab:2}. For the most optimistic outcome, the contribution to the observed neutrino background is $\lesssim 22\%$.

\begin{deluxetable*}{c|ccccc|c}[htpb!] \label{Tab:2}
	\tablecaption{Theoretical Event Rate Densities and Diffuse Fractional Fluxes for AGN Disk Transients}
	\tablecolumns{7}
	
	\setlength{\tabcolsep}{5mm}{}
	
	\tablewidth{0pt}
	\tablehead{		\colhead{Explosion} &\colhead{$M_{\bullet}$} &\colhead{$r/r_{\rm S}$} &\colhead{$R_0/\mathrm{Gpc}^{-3}\mathrm{yr}^{-1}$} &\colhead{$\rho_{c.-15}$} &\colhead{$\alpha$} &\colhead{Diffuse Fractional Flux} }
	\startdata
	SN Ia    & $10^7$ & $10^3$ & \textless5000 & 100 & 1.5 & $\lesssim22.3\%$\\
	SN Ia    & $10^7$ & $10^3$ & \textless5000 & 10  & 1.5 & $\lesssim0.9\%$\\
	SN Ia    & $10^7$ & $10^3$ & \textless5000 & 1   & 1.5 & $\lesssim0.07\%$\\
	SN Ia    & $10^7$ & $10^3$ & \textless5000 & 100 & 2 & $\lesssim5\%$\\
	SN Ia    & $10^7$ & $10^3$ & \textless5000 & 10  & 2 & $\lesssim0.19\%$\\
	SN Ia    & $10^7$ & $10^3$ & \textless5000 & 1   & 2 & $\lesssim0.062\%$\\
	SN Ia    & $10^7$ & $10^3$ & \textless5000 & 100 & 3 & $\lesssim0.48\%$\\
	SN Ia    & $10^7$ & $10^3$ & \textless5000 & 10  & 3 & $\lesssim0.080\%$\\
	SN Ia    & $10^7$ & $10^3$ & \textless5000 & 1   & 3 & $\lesssim0.061\%$\\
	\hline
	SN Ia (SS)    & $10^7$ & $10^3$ & \textless5000 & 1   & -- & $\lesssim7.23\%$\\
	\hline
	Kilonova & $10^7$ & $10^3$ & \textless460  & 10  & 2 & $\lesssim0.019\%$\\ 
	CCSN     & $10^7$ & $10^3$ & \textless100  & 10  & 2 & $\lesssim0.008\%$\\
	GRB-SN   & $10^7$ & $10^3$ & \textless1    & 10  & 2 & $\lesssim0.0006\%$\\ 
	\enddata
	\tablecomments{We assume all binary WD mergers can produce SNe Ia, while all binary NS mergers and  $\sim20\%$ NS–BH mergers can power kilonovae \citep{mckernan2020black,zhu2021NSBH}. 
	The local event rate density of AGN CCSN is based on the constraint by \cite{grishin2021supernova}. The rate densities of AGN GRB-SNe are assumed to be $\sim1\%$ of AGN CCSNe \citep{dittmann2021accretion}}
	
\end{deluxetable*}

The realistic contribution of AGN disk transients to the diffuse neutrino background depends on the position distribution of these transients, i.e., the radial distance $r$ and the vertical height $h$. However, the accurate location of these transients in the AGN disk is still quite unclear. For the vertical height $h$, AGN disk transients are likely to take place at the height with the higher star forming rate (SFR), namely, higher gas density. For a disk with the Gaussian density profile, AGN disk transients may tend to occur around the mid-plane of the disk where the gas density is highest in the vertical direction and a large position deviation from the mid-plane will lead the gas density to decrease remarkably for a Gaussian density profile. Therefore, our assumption that AGN disk transients occur at the mid-plane of the disk may be reasonable. For the radial distance $r$, we can evaluate the radial dependence of AGN disk transients based on the well-known Kennicutt-Schmidt (K-S) relation \citep{1998ApJ...498..541K} which is usually used to estimate the correlation between the SFR surface density and the gas surface density. The K-S relation can be written as ${\Sigma _{\rm SFR}} = A\Sigma _{\rm gas}^n$, where $n\simeq 1.14$ is the power relation index \citep{2015A&A...577A.135C,2023ApJ...944..159F}, ${\Sigma _{\rm SFR}}$ is the SFR surface density, $\Sigma _{\rm gas}$ is the gas surface density, and $A$ is the normalization constant representing the efficiency of the processes regulating gas-stars conversion. Here, the gas surface density is ${\Sigma _{gas}} \propto {\rho _0}(r)H \propto {r^{ - 3/2}}$ and then, one gets ${\Sigma _{SFR}} \propto {r^{ - 1.71}}$ for $n= 1.14$. Thus, the radial dependence of the SFR can be evaluated as ${R_{SFR}} \propto {r^{0.29}}$, which depends slightly on the radius. For the radius in the range of $10^3-10^5\,r_{\rm S}$, the difference of the SFR is no more than a factor of $4$. For the lower density of the circum-disk materials, AGN disk transients occurring at the larger radii will produce higher neutrino fluences since more disk materials will consume the shock kinetic energy \citep{zhu2021high2}, which induces the actual contribution of AGN disk transients to the diffuse neutrino background would be higher than our estimations that base on the radius $r=10^3 r_{\rm S}$. For the high enough density of circum-disk materials, the shock kinetic energy will be almost exhausted by the AGN disk and the circum-disk materials whether for the small radius or the large radius, so in this situation, the radial dependence will be weak. Besides, the SMBH mass will affect the gas distribution of the disk, especially the mid-plane gas density $\rho_0$, and consequently affect the neutrino production. However, as indicated by the SMBH mass function \citep{2004MNRAS.354.1020S,2019MNRAS.487..198G}, the SMBH mass distribution index ranging $10^6-10^8\,M_{\odot}$ is relatively flat (we ignore the relatively small number of SMBH with the mass larger than $10^8\,M_{\odot}$), and therefore adopting a middle value of the SMBH mass, i.e., $10^7\,M_{\odot}$, can eliminate the effects of uncertainties of SMBH masses to some extent. As a result, although these parameters such as positions of AGN transient occurring at AGN disks and the precise SMBH mass function have uncertainties, our estimations locates at a reasonable range and may not deviate from the realistic situation significantly.

\section{Discussion}\label{sec:discusssion}

In this paper, we investigate neutrino production by transient explosions in the AGN accretion disk. By considering the AGN outflow and disk wind, the circum-disk materials become very extended compared with the general Gaussian density profile of a stable gas-dominated disk. Due to the existence of circum-disk gases, the shock kinetic energy will be further consumed and eventually contribute more neutrino production. We present the lightcurves of neutrino emission after the shock breakout and obtain a longer neutrino signal duration due to the extended circum-disk gases. Consequently, we determine that the neutrino fluence for a single source will be enhanced significantly, inducing a much higher contribution to the diffuse neutrino background as well. Neutrino detection for a single source could be expected within a distance of a few Mpc and the contribution to the diffuse neutrino background can reach $\sim 20\%$ for an optimistic circum-disk gas environment. The precise neutrino production depends on the explosion energy, the ejecta mass, the radial and height distribution of the source, as well as the gas distribution on the disk and in the circum-disk, which has been explored in the paper. Basically, the neutrino production will be enhanced significantly due to the extra materials surrounding AGN disks compared with the disk atmosphere with a pure Gaussian distribution as adopted in \cite{zhu2021high}.

In addition to SNe Ia, kilonovae, CCSNe, and GRB-SNe, other potential sources, e.g., Bondi explosions of stellar-mass BHs \citep{wang2021accretion1}, and micro-tidal disruption events (mTDEs) \citep{yang2022}, may also take place in AGN accretion disks. These energetic events are predicted to have high event rate densities in AGN disks and may contribute a considerable fraction of the diffuse neutrino background. Recently, a few potential AGN disk transients were detected, although their origins were still unclear \citep{holoien2022,hinkle2022}, and more AGN disk transients might be recorded in previous optical survey projects. Lightcurves of AGN disk transients could have similar peak brightness, peak time, and evolution pattern with those of TDEs \citep{zhu2021thermonuclear,grishin2021supernova,ren2022} so that some AGN disk transients could be wrongly identified as TDEs. Future theoretical studies and observations may extend our understanding and help to give a better estimation of the contribution to the neutrino background by AGN disk transients.

In recent, IceCube performed a neutrino follow-up search on the candidate optical counterpart to the BBH event GW190521, which is thought to possibly occur in the AGN disk~\citep{2023ApJ...944...80A}. No significant neutrino emission was observed and consequently, a $90\%$ upper limit on the neutrino fluence ($\epsilon_{\nu}^2 \phi_{\nu}$) of $\sim 0.05\,\rm GeV \,cm^{-2}$ is derived. The relatively far luminosity distance of $3931\pm953\,\rm Mpc$ \citep{graham2020} made the fluence of the possible neutrino production low and difficult to be detected by a single source. In addition, the dynamic evolution of the BBH event that may launch relativistic jets via the hyper-Eddington accretion of the remnant BH \citep{wang2021accretion2,2023arXiv230107111T} unlike the isotropic ejecta of SN Ia could be different in the AGN disk, although the interaction between the relativistic jet and AGN disk could produce the high-energy neutrino emission more efficiently through $p\gamma$ interactions as well \citep{zhu2021high2}. The increasing rate of expected compact object mergers in the AGN disk can be expected with the upcoming O4 run of LIGO and Virgo, the search for the association between gravitational wave events and high-energy neutrinos would provide more opportunities to conduct multi-messenger studies.

AGN disk transients could be the natural \textit{hidden cosmic-ray accelerators} that generate a higher ratio between neutrino and gamma-ray production than the general predictions of hadronic processes, which can alleviate the tension between the diffuse neutrino background and the isotropic diffuse gamma-ray background to some extent \citep{2016PhRvL.116g1101M}. In the dense environment of the AGN disk and the circum-disk materials, the EM radiation of these transients, the disk radiation, or even the corona radiation could attenuate the gamma-rays significantly. Eventually, a much higher neutrino flux than the gamma-ray flux can be expected, which is similar to the observational features of the neutrino outburst from the blazar TXS 0506+056 \citep{2018Sci...361..147I} and the recent neutrino observations from the nearby active galaxy NGC 1068 \citep{2022Sci...378..538I}. Further studies can help us to clarify the populations of neutrino sources.

\acknowledgments

We thank Prof. Bing Zhang for his valuable comments. We also thank the anonymous referee for his/her detailed and helpful comments, which have helped us to improve this paper. This work is supported by National Natural Science Foundation of China under grants No.12003007 and the Fundamental Research Funds for the Central Universities (No. 2020kfyXJJS039).

\bibliography{reference}{}
\bibliographystyle{aasjournal}
\end{document}